\documentclass[twocolumn,prl,bibnotes,showpacs,superscriptaddress]{revtex4-1}

\usepackage{amsmath,amsfonts,amssymb}
\usepackage{bm}
\usepackage{dcolumn}
\usepackage[mathscr]{eucal}
\usepackage[dvips]{graphicx,graphics,color,epsfig}
\usepackage{latexsym}
\usepackage{natbib}
\usepackage{verbatim}
\usepackage[colorlinks=true,linkcolor=blue,citecolor=blue,urlcolor=blue,breaklinks=true]{hyperref}
\usepackage{bm}
\usepackage{tabulary}

\renewcommand{\vec}[1]{\mbox{\boldmath$#1$}}

\begin{document}

\title{Plasmon-enhanced second harmonic generation in the laser-irradiated
cubic metal nanoparticles}
\author{S.~V. Fomichev}
\email{fomichev\_sv@nrcki.ru}
\affiliation{National Research Center ``Kurchatov Institute", 1 Kurchatov
place, 123182 Moscow, Russia}
\affiliation{Moscow Institute of Physics and Technology, 141700 Dolgoprudny,
Moscow Region, Russia}
\author{A.~M. Bratkovsky}
\email{alex.bratkovski@gmail.com}
\affiliation{Corning Science and Technology, Corning, New York 14831}
\affiliation{P.~L. Kapitza Institute for Physical Problems, 2 Kosygina Str.,
119334 Moscow, Russia}

\begin{abstract}
The plasmon-enhanced second harmonic generation in the subwavelength neutral
metal \emph{cubic} nanoparticles is calculated for the first time in the
hydrodynamic and cold plasma approximations. The theory is developed that
takes into account all singularities of the electromagnetic field at the
cubic nanoparticle surface. The results are compared to the linear case for
the nanocube and to those for a spherical nanoparticle. In the latter case,
they demonstrate very strong enhancement of the local field strength and the
second harmonic signal.
\end{abstract}

\maketitle

\textit{Introduction}. -- The nonlinear optical properties of subwavelength
metal nanoparticles and, in particular, the second harmonic generation are
attracting a rapidly growing interest. A lot of effort was directed to the
case of spherical and spheroidal nanoparticles, both theoretically \cite%
{Gersten,Liebsch,Hayata,Ostling,Dewitz,Dadap1,Dadap2,Markel,Huo,Tomchuk,Butet2}
and experimentally \cite%
{Antoine,Brevet1,ZR,Viarbitskaya,Butet1,Brevet2,Capretti}, while much less
attention was paid so far to the case of non-spherical nanoparticles,
perhaps because of complexity of the problem \cite%
{Neacsu,Chan,Zeng,Scarola,Gusakov15}. For example, a lot of papers (see
\cite{FBr} and references therein) are devoted to studying \textit{linear}
plasmons in cubic metal nanoparticles, but there is no corresponding
investigations of the nonlinear optical properties of cubic nanoparticles
and, in particular, the laser induced plasmon-assisted second harmonic
generation (SHG). It is worth mentioning a thrust in using high-Z (Sn)
plasma \cite{EUV} and the charged metal particles \cite{Krainov02,Gusakov15}
for high-harmonic generation that may be used in EUV lithography and other
applications by way of stripping electrons by laser with their subsequent
recapture by the charged ion or the metal particle. Here, we are not
interested in multiple ionization regime and will assume that the particles
remain neutral at all times. In practice, this corresponds to laser power
below some 10~GW/cm$^2$ \cite{Gusakov15}. The purpose of this work is to
investigate the SHG in ideal cubic nanoparticles with free electrons both in
far-field and in the near-field and to compare it with the spherical case.

\textit{Model}. -- The basic method for calculating the low order harmonic
generation in the subwavelength metal nanoparticles of arbitrary shape was
published in Refs.~\cite{FBr,FB1,FB2}. Using the perturbation theory with
respect to the incident laser field \cite{FBr,FB1,FB2}, we start from the
dimensionless equations for the first-order corrections to the respective
quantities defining the alternating electromagnetic field inside the
subwavelength neutral metal nanoparticle within the charge compensation
approximation (CCA) where the static (zeroth-order) component of the
electric field inside the nanoparticle including its narrow boundary
vanishes identically ($\vec{\mathcal{E}}_0\equiv 0$). Introducing the
dimensionless coordinates as $\boldsymbol{\rho}=\vec{r}/a$ with $a$ the
cube edge length, and reducing the electron density by dividing it by
$z_{i}n_{\mathrm{ion}}$, where $z_{i}$ is the mean ionic charge and
$n_{\mathrm{ion}}$ is the reference bulk ion density of the nanoparticle
substance, these equations read as
\begin{subequations}
\begin{equation}
\frac{i(\mathscr{N}_0-\widetilde{\omega}^2-i\gamma\widetilde{\omega})
\vec{\mathcal{E}}_1+i\mathscr{N}_0\vec{\mathcal{E}}_L}{\widetilde{\omega}^2
+i\gamma\widetilde{\omega}}=\mathrm{curl}\,\vec{h}_1\,,
\label{eq1a}
\end{equation}
\begin{equation}
\vec{\mathcal{E}}_1=-\nabla\varphi_1\,, \quad \mathrm{div}\,\vec{h}_1=0\,.
\label{eq1b}
\end{equation}
\end{subequations}
The expression $\mathscr{N}_1=\nabla^2\varphi_1$ determines the linear
correction to the electron density, while $\mathscr{N}_0$ defines the static
(equilibrium) electron density distribution in the nanoparticle when the laser
field is off. Equations~(\ref{eq1a})--(\ref{eq1b}) are the four first-order
differential equations for four quantities, that is for three components of
the magnetic vector function $\vec{h}_1$ and the scalar potential $\varphi_1$.
In these equations, $\widetilde{\omega}=\omega/\omega_p$ is the reduced laser
frequency with respect to the bulk plasma frequency $\omega_p=
\sqrt{4\pi e^{2}z_{i}n_{\mathrm{ion}}/m_e}$ with $m_e$ and $e$ the electron
mass and charge, respectively, $\gamma$ the dimensionless phenomenological
damping constant and $\vec{\mathcal{E}}_L=\vec{E}_L/E_0$ the reduced laser
electric field amplitude, with $E_0=4\pi ez_{i}n_{\mathrm{ion}}a$. Note that
$E_0\sim 10^{10}$~V/cm at typical values $z_{i}\sim 1$,
$n_{\mathrm{ion}}\sim 10^{22}$~cm$^{-3}$ and $a\sim 10$~nm, so in our
conditions $\mathcal{E}_L$ is always extremely small with respect to unity.
All components of the induced electromagnetic field inside the nanoparticle,
the harmonic amplitudes $\boldsymbol{\mathcal{E}}_n$ and
$\boldsymbol{\mathcal{H}}_n$ with $n=1,2\dots$ are normalized by the same way.
Besides, magnetic vector function $\vec{h}_n$ is defined as $\vec{h}_n=
\boldsymbol{\mathcal{H}}_n/(\omega a/c)$, with $c$ the speed of light and
$\omega a/c$ the dipole approximation applicability parameter that should be
much less than unity for the subwavelength nanoparticles. In the above
equations, the induced magnetic field inside the nanoparticle that enters
through the magnetic vector function neither can be eliminated nor neglected,
so even in the dipole approximation for subwavelength nanoparticles this can
not be reduced to the standard case of the electrostatic approximation.

The case of \emph{linear} scattering and absorption of the laser field by
cubic and parallelepiped-like nanoparticles in the framework of the developed
model was considered in detail in Ref.~\cite{FBr}. The solution of the
first-order equations should be used as the source term in the second-order
equations, which are written as
\begin{subequations}
\begin{align}
&\frac{i(\mathscr{N}_0-4\widetilde{\omega}^2-2i\gamma\widetilde{\omega})
\vec{\mathcal{E}}_2}{\left(2\widetilde{\omega}^2+i\gamma\widetilde{\omega}
\right)}=\frac{i\widetilde{\omega}^2\vec{V}_2}{\mathscr{N}_{0}^{{\kern1pt}2}
\left(2\widetilde{\omega}^2+i\gamma\widetilde{\omega}\right)}+
\mathrm{curl}\,\vec{h}_2\,, \label{eq2a} \\
&\vec{\mathcal{E}}_2=-\mbox{\boldmath{$\nabla$}}\varphi_2\,, \quad
\mathrm{div}\,\vec{h}_2=0\,, \label{eq2b}
\end{align}
\end{subequations}
with
\begin{align}
&V_{2\alpha}=-\left(-i+\frac{\gamma}{\widetilde{\omega}}\right)
2q_{1\alpha}\mathscr{N}_0\mathscr{N}_1-\mathscr{N}_0\frac{\partial
(q_{1\alpha}q_{1\beta})}{\partial\rho_{\beta}} \notag \\
&+q_{1\alpha}q_{1\beta}\frac{\partial\mathscr{N}_0}{\partial\rho_{\beta}}
-3\frac{\mathcal{E}_{0\alpha}}
{\widetilde{\omega}^2}\mathscr{N}_0\mathscr{N}_{1}^{{\kern1pt}2}
-3\frac{\mathcal{E}_{1\alpha}+\mathcal{E}_{{L}{\alpha}}}{\widetilde{\omega}^2}
\mathscr{N}_{0}^{{\kern1pt}2}\mathscr{N}_1 \label{eq3}
\end{align}
and
\begin{equation}
\vec{q}_1=\frac{\mathscr{N}_0\left(\vec{\mathcal{E}}_1+\vec{\mathcal{E}}_L
\right)}{i\left(\widetilde{\omega}^2+i\gamma\widetilde{\omega}\right)}\,.
\label{eq4}
\end{equation}
Here, the equation $\nabla^2\varphi_2=\mathscr{N}_2$ holds for the
second-order correction to the electron density.

Generally, the boundary conditions to these equations read as
\begin{align}
&\frac{1}{4\pi \epsilon_{n}}\oint d\boldsymbol{S}^{\prime}\left\{\!\left[
\varphi_{n}(\boldsymbol{\rho}^{\prime})-\varphi_{n}(\boldsymbol{\rho})
\right]\frac{\boldsymbol{\rho}-\boldsymbol{\rho}^{\prime}}
{|\boldsymbol{\rho}-\boldsymbol{\rho}^{\prime}|^3}
-\frac{\nabla\varphi_{n}(\vec{\rho}^{\prime})}{|\boldsymbol{\rho}
-\boldsymbol{\rho}^{\prime}|}\right\} \notag \\
&=\varphi_{n}(\boldsymbol{\rho})-\delta_{1n}\frac{\epsilon_{1}-1}
{\epsilon_1+2}(\vec{\mathcal{E}}_{L}\cdot\boldsymbol{\rho}) \label{eq5}
\end{align}
and
\begin{align}
&\vec{h}_{n}(\boldsymbol{\rho})+\frac{1}{4\pi}\oint \frac{\left[
(\boldsymbol{\rho}-\boldsymbol{\rho}^{\prime})\times\left[
d\boldsymbol{S}^{\prime}\times\left(\vec{h}_{n}(\boldsymbol{\rho}^{\prime})
-\vec{h}_{n}(\boldsymbol{\rho})\right)\right]\right]}{|\boldsymbol{\rho}
-\boldsymbol{\rho}^{\prime}|^{3}} \notag \\
&-\oint\frac{\left[d\vec{S}^{\prime}\times\mathrm{curl}\,\vec{h}_{n}
(\vec{\rho}^{\prime})\right]}{4\pi|\boldsymbol{\rho}
-\boldsymbol{\rho}^{\prime}|}-\frac{1}{4\pi}\!\int\!d^{{\kern1pt}3}\!
\rho^{\prime}\left[h_{n\beta}(\boldsymbol{\rho}^{\prime})
-h_{n\beta}(\boldsymbol{\rho})\right] \notag \\
&\times\frac{3(\rho-\rho^{\prime})_{\alpha}(\rho-\rho^{\prime})_{\beta}
-\delta_{\alpha\beta}|\boldsymbol{\rho}-\boldsymbol{\rho}^{\prime}|^{2}}
{|\boldsymbol{\rho}-\boldsymbol{\rho}^{\prime}|^{5}}=0\,, \label{eq6}
\end{align}
where $n$ is the order of nonlinearity and $\epsilon_{n}=\epsilon(n\omega)$
is the dielectric permittivity of the surrounding medium on the $n$th-order
laser harmonic frequency. However, in this specific form these boundary
conditions are applicable only to the nanoparticles with diffuse surface, when
all the quantities including the ion density vary smoothly at the narrow
nanoparticle surface. This program was realized for the calculation of the
nonlinear properties of metallic spherical nanoparticles in
Refs.~\onlinecite{FB1,FB2}. But in the opposite limiting case of the sharp
nanoparticle boundary, which we consider here for the cubic nanoparticles, the
step-like static electron density $\mathscr{N}_0(\vec{\rho})$ is equal to one
or zero inside or outside the nanoparticle, respectively. In this case,
additional boundary conditions following immediately from Eqs.~(\ref{eq1a})
and (\ref{eq1b}) or Eqs.~(\ref{eq2a}) and (\ref{eq2b}), respectively, for
$n=1$ or $n=2$ in our case should be taken into consideration to allow for the
step-like and higher singularities in the quantities $\nabla\varphi_n$ and
$\mathrm{curl}\,\vec{h}_n$ at the sharp nanoparticle surface. In the linear
case, it follows from Eqs.~(\ref{eq1a}) and (\ref{eq1b}) that due to steplike
discontinuity of the static electron density $\mathscr{N}_0$ at the sharp
nanoparticle surface, the steplike discontinuities should be also present
for some components of $\vec{\mathcal{E}}_1$ and $\mathrm{curl}\,\vec{h}_1$,
while $\varphi_1$ and $\vec{h}_1$ are themselves continuous. Here, the
first-order electron density correction $\mathscr{N}_1$ has the
$\delta$-function type singularity at the nanoparticle surface. On the other
hand, the steplike discontinuities for some components of
$\vec{\mathcal{E}}_1$ and the singularity of $\mathscr{N}_1$ generate the
$\delta$-function type singularity in function $\vec{V}_{\!2}$, which plays
the role of the source term in the second-order equations~(\ref{eq2a}) and
(\ref{eq2b}). In turn, it generates the $\delta$-function type singularities
in some components of $\vec{\mathcal{E}}_2$ and $\mathrm{curl}\,\vec{h}_2$,
and, as a consequence, the steplike discontinuities in $\varphi_2$ and some
components of $\vec{h}_2$. All this should be taken into account in the
calculations with the details to be described elsewhere.

The second-order near electric field can be calculated through the
second-order electric potential outside the nanoparticle,
\begin{align}
&\varphi_{n}(\boldsymbol{\rho})=-\frac{1}{4\pi\epsilon_{n}}\!\int
\!d^{{\kern1pt}3}\!\rho^{\prime}\frac{\mathscr{N}_n
(\boldsymbol{\rho}^{\prime})}{|\boldsymbol{\rho}-\boldsymbol{\rho}^{\prime}|}
=-\frac{1}{4\pi\epsilon_{n}}\!\int\!d^{{\kern1pt}3}\!\rho^{\prime}
\frac{\nabla^2\varphi_n(\boldsymbol{\rho}^{\prime})}{|\boldsymbol{\rho}
-\boldsymbol{\rho}^{\prime}|} \notag \\
&=\frac{1}{4\pi\epsilon_{n}}\oint d\boldsymbol{S}^{\prime}\left(\varphi_{n}
\frac{\boldsymbol{\rho}-\boldsymbol{\rho}^{\prime}}{|\boldsymbol{\rho}
-\boldsymbol{\rho}^{\prime}|^3}-\frac{1}{|\boldsymbol{\rho}
-\boldsymbol{\rho}^{\prime}|}\boldsymbol{\nabla}\varphi_n\right),
\label{eq7}
\end{align}
as $\vec{\mathcal{E}}_2=-\mbox{\boldmath{$\nabla$}}\varphi_2$.

For the far-field calculations, the nanoparticle dipole moment for the
$n$th-order harmonic can be calculated as
\begin{align}
&d_{n\alpha}=-\int\rho_{\alpha}\,\mathscr{N}_n\frac{d^{{\kern1pt}3}\!\rho}
{4\pi}=-\int\rho_{\alpha}\,\nabla^2\varphi_n\frac{d^{{\kern1pt}3}\!\rho}{4\pi}
\notag \\
&=\frac{1}{4\pi}\oint\left\{\varphi_{n}\,dS_{\alpha}-\rho_{\alpha}
(d\boldsymbol{S}\cdot\boldsymbol{\nabla})\varphi_n\right\}, \label{eq8}
\end{align}
where the integration is taken over the outer nanoparticle boundary. The
intensity $I_{n}$ of the radiation scattered due to the induced dipole at the
$n$th-order harmonic frequency $n\omega$ is defined by the corresponding
dipole moment and is given as
\begin{equation}
I_{n}\sim(n\omega)^{4}|\vec{d}_{n}|^{2}. \label{eq9}
\end{equation}
However, it defines both the scattering and the absorbing properties of the
cubic nanoparticles only for odd harmonics including the linear case at $n=1$,
due to the symmetry properties. For the second harmonic, the dipole scattering
intensity vanishes, and it can be used only for testing the calculation code.
Thus, for the second-harmonic the scattered intensity is defined by the
induced \emph{quadrupole} moment of the nanoparticle. The dimensionless
second-order quadrupole tensor $Q_{2\alpha\beta}$, which is defined in direct
analogy with Eq.~(\ref{eq8}) for the $n$th-order dipole moment, is given by
\begin{align}
Q_{2\alpha\beta} & =-\int(3\rho_{\alpha}\rho_{\beta}-\rho^{2}
\delta_{\alpha\beta})\mathscr{N}_2\,\frac{d^{{\kern1pt}3}\!\rho}{4\pi}
\notag \\
&=-\int(3\rho_{\alpha}\rho_{\beta}-\rho^2\delta_{\alpha\beta})
\nabla^2\varphi_2\,\frac{d^{{\kern1pt}3}\!\rho}{4\pi} \notag \\
&=\frac{1}{4\pi}\oint\{\varphi_2[3\rho_{\alpha}dS_{\beta}+3\rho_{\beta}
dS_{\alpha}-2(\boldsymbol{\rho}\cdot d\boldsymbol{S})\delta_{\alpha\beta}]
\notag \\
&-(3\rho_{\alpha}\rho_{\beta}-\rho^2\delta_{\alpha\beta})(d\boldsymbol{S}
\cdot\boldsymbol{\nabla})\varphi_2\}\,. \label{eq10}
\end{align}
The mean power $P_{2}^{(Q)}$ of quadrupole second-harmonic generation is
defined by the dimensionless quadrupole tensor $Q_{2\alpha\beta}$ and is
given as
\begin{equation}
P_{2}^{(Q)}\sim(2\omega)^{6}|Q_{2\alpha\beta}|^2\,, \label{eq11}
\end{equation}
with the quadratic dependence on the incident laser wave intensity. The
right-hand side of Eq.~(\ref{eq11}) completely determines the frequency
dependence of SHG. The results for these dependencies will be presented next.

\begin{figure*}[tbp]
\centering
\includegraphics[]{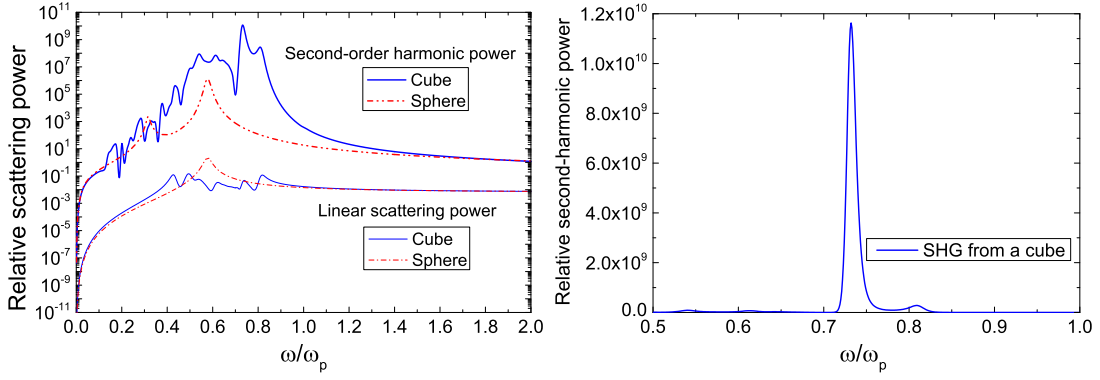}
\caption{(Color online) The left panel: the laser frequency dependence in
the log-scale of the normalized linear scattering power and the second
harmonic power for ideal cubic nanoparticle in vacuum (thick and thin solid
curves, respectively) as compared with a spherical one of the same volume
(dash-double-dotted and dash-dotted curves). The right panel: the frequency
dependence of the second harmonic power for the cube in vacuum in the normal
scale. For the cube, the electric field amplitude of the incident
linear-polarized laser wave is directed along the cube axis perpendicularly
to its facets. The damping constant is $\protect\gamma=0.03$.}
\label{Fig:1}
\end{figure*}
\begin{figure*}[tbp]
\centering
\includegraphics[]{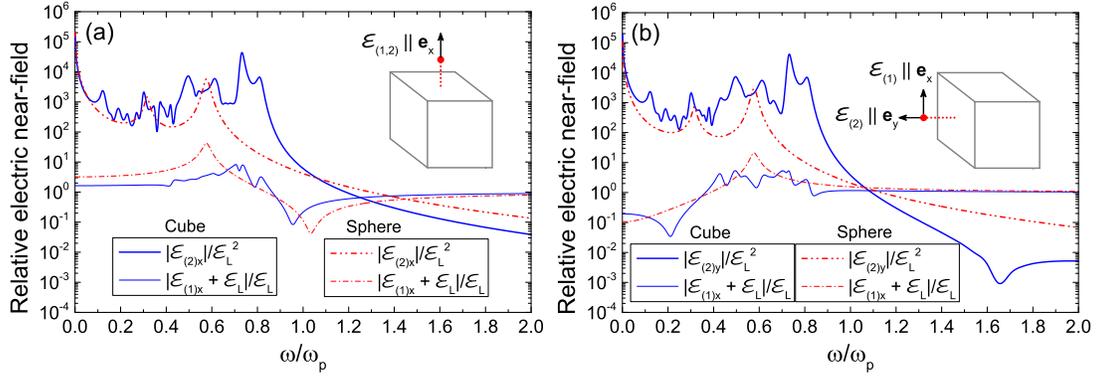}
\caption{(Color online) The left panel: the laser frequency dependence of
the normalized amplitudes of the total electric near-field components
corresponding to the linear response and the second harmonic generation for
ideal cubic nanoparticle in vacuum (thick and thin solid curves,
respectively) just above the center of the top cube face (at
$\protect\rho_x=0.6$, $\protect\rho_y=\protect\rho_z=0$) oriented
perpendicularly to the laser electric field directed vertically along the
$x$-axis, as compared with the same for a spherical nanoparticle of the same
volume (dash-double-dotted and dash-dotted curves). The right panel: the same
just to the left of the center of the left cube side (at
$\protect\rho_y=-0.6$, $\protect\rho_x=\protect\rho_z=0$) oriented in parallel
to the laser electric field directed vertically along the $x$-axis. The
damping constant is $\protect\gamma=0.03$.}
\label{Fig:2}
\end{figure*}

\textit{Numerical results}. -- The calculations for the cubic nanoparticles
have been performed on the three-dimensional uniform grid with up to
$42\times 42\times 42$ nodal points to get a fair convergence for the second
harmonic case, and a good convergence for the linear case, with the whole
cube located at $-0.5\leq \rho_x,\rho_y,\rho_z \leq 0.5$ in the reduced
coordinate variables. In this case, the first-order electric potential and
three magnetic field components in the cube bulk were given on the cubic grid
points themselves, on the vertices of the cubic grid unit cells, while three
\textit{outside} surface electric field components were given on the centers
of the cubic grid cell faces on the whole cube \emph{surface}. All these
quantities constitute the complete set of first-order variables, which satisfy
a closed set of field equations together with the boundary conditions for a
cube. The second-order variables are defined similarly, but in this case the
number of variables somewhat increases due to necessity to introduce two sets
of the surface variables for the second-order electric potential and some
magnetic field components due to their discontinuity on the steplike surface
of the nanoparticle.

\begin{figure*}[tbp]
\includegraphics[]{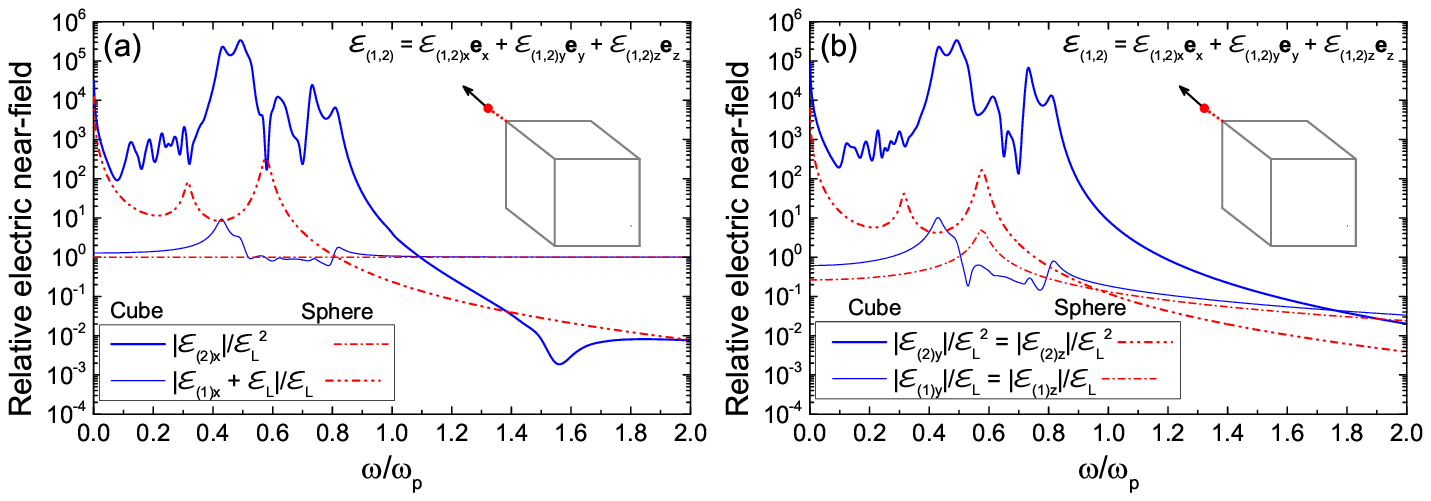}
\caption{(Color online) The left panel: the laser frequency dependence of
the normalized amplitudes of the $x$-component of the total electric
near-field corresponding to the linear response and the second harmonic
generation for ideal cubic nanoparticle in vacuum (thick and thin solid
curves, respectively) just near the vertex of the cube along the cube
diagonal (at $\protect\rho_x=\protect\rho_z=0.558$, $\protect\rho_y=-0.558$),
with the laser electric field directed vertically along the $x$-axis, as
compared with the same for a spherical nanoparticle of the same volume
(dash-double-dotted and dash-dotted curves). The right panel: the same for the
normalized amplitudes of the $y$- and $z$-components of the total electric
near-field corresponding to the linear response and the second harmonic
generation, with the laser electric field directed vertically along the
$x$-axis. The damping constant is $\protect\gamma=0.03$.}
\label{Fig:3}
\end{figure*}

The calculations were performed for both the far-field and the near-field
output quantities for the second harmonic. We present the results for the
far-field in the vacuum in Fig.~\ref{Fig:1}, while the results for the
electric near-field are given in Fig.~\ref{Fig:2} and Fig.~\ref{Fig:3}. The
power of the second harmonic scattered by the nanoparticles versus the laser
frequency is presented in the normal scale in the right panel of
Fig.~\ref{Fig:1}, with the dominant resonance peak located at
$\omega\approx 0.73\omega_p$. In the left panel of Fig.~\ref{Fig:1}, we
present the detailed structure of the frequency dependence of the scattered
second-harmonic power, together with the corresponding results for the
linear light scattering by the cubic nanoparticles obtained in the same
calculations. Also, the results for both the linear scattering and the
second harmonic generation in the spherical nanoparticle of the same volume
are shown in this figure for comparison.

First of all, let us note that the asymptotic behavior of the second harmonic
power for an ideal cube and for a sphere of the same volume coincide both at
low and at high laser frequencies (with respect to the Mie plasmon frequency
in a sphere at $\omega/\omega_p\approx 0.577$ that corresponds to the
strongest peak on the dash-dotted curves). Then, the SHG power in an ideal
cube is generally much larger than in a sphere for the mid-range frequencies,
especially in the range of the spherical Mie plasmon resonance with a broad
maximum on a logarithmic scale shifted to the higher frequencies, where we
observe the dominant narrow resonance peak at $\omega\approx 0.73\omega_p$. It
is in variance with a linear case, where the linear Mie plasmon resonance in a
sphere dominates over scattering power in a cube. Note also that the dominant
resonance peak in the SHG power in a cube corresponds to some resonance peak
in the linear light scattering, which is a relatively weak and does not belong
to three largest peaks in the linear light scattering in a cube \cite{FBr}. On
the other hand, the SHG powers in a sphere and in a cube are comparable in the
range of the quadrupole resonance in SHG in a sphere that corresponds to the
left peak on the dash-double-dotted curve in the left panel of
Fig.~\ref{Fig:1}, while it exhibits more complex resonance structure.

We present the results on laser frequency dependence of the normalized (not
dependent on the incident laser field) amplitudes of the total electric
near-field components corresponding to both the linear response [for
$(\boldsymbol{\mathcal{E}}_1+\boldsymbol{\mathcal{E}}_L)/\mathcal{E}_L$]
and the second harmonic generation [for
$\boldsymbol{\mathcal{E}}_2/\mathcal{E}_{L}^{2}$] for ideal cubic
nanoparticles in vacuum (thick and thin solid curves, respectively, in
Figure~\ref{Fig:2}) at the point just above the center of the top cube face
(at $\rho_x=0.6$, $\rho_y=\rho_z=0$, the left panel), and just to the left of
the center of the left cube side (at $\rho_y=-0.6$, $\rho_x=\rho_z=0$, the
right panel). The laser electric field is assumed to be directed vertically,
along the $x$-axis. For comparison, similar results for spherical
nanoparticles of the same volume (dash-double-dotted and dash-dotted curves)
are also presented in both panels of Fig.~\ref{Fig:2}. Note first of all that
if in the first case (the left panel of Fig.~\ref{Fig:2}) both the linear
response field and the second-harmonic field are directed along the $x$-axis
(the direction of the laser electric field), in the second case (the right
panel of Fig.~\ref{Fig:2}) the electric field of the second harmonic is
directed perpendicular to it and the corresponding cube face, while the linear
response field is still parallel to the laser electric field. This fact
validates the dipole character of the linear response field and the
\emph{quadrupole} character of the field of the second harmonic. Secondly,
note that the laser frequency dependencies of the second harmonic electric
field in both cases (the left and the right panels of Fig.~\ref{Fig:2}) are
quite similar, with the dominant resonance peak at
$\omega\approx 0.73\omega_p$, as was the case for the far-field second
harmonic power in Fig.~\ref{Fig:1}. Generally, the second harmonic field near
the cube is higher than near the sphere although it is not the case for high
frequency asymptotic range.

In Figure~\ref{Fig:3}, the laser frequency dependencies of the normalized
amplitudes of total electric near-field components corresponding to the
linear response and the SHG are presented for ideal cubic nanoparticles in
vacuum (thick and thin solid curves, respectively) just near the vertex of
the cube along the cube diagonal (at $\rho_x=\rho_z=0.558$, $\rho_y=-0.558$). 
The laser electric field is directed vertically along the $x$-axis, and the 
comparison with the same for spherical nanoparticles of the same volume 
(dash-double-dotted and dash-dotted curves) is given again. The results for 
the $x$-components are given in the left panel of Fig.~\ref{Fig:3}, while the 
right panel exhibits the results for $y$- and $z$-components of the total 
electric near-field corresponding to the linear response and the second 
harmonic generation. The main difference of this case from the previous one in 
Fig.~\ref{Fig:2} is the near-field resonance with two dominant peaks 
(originating from double-split broad maxima) located at 
$\omega/\omega_p\approx 0.43$ and $\omega/\omega_p\approx 0.49$. Note that 
those peaks are also present in the far-field in Fig.~\ref{Fig:1} and in the 
near-field cases, Fig.~\ref{Fig:2}, but they are not dominant. On the other 
hand, the resonance at $\omega\approx 0.73\omega_p$ is also present in the 
case of Fig.~\ref{Fig:3}, but not as the dominant one. Besides, it should be 
noted that in this case there is much more noticeable difference between the 
cube and the sphere with strong increase of the second harmonic field near the 
cube with respect to that near the sphere of the same volume. It can be
connected with the point position near the cube vertex chosen for the results 
presented in Fig.~\ref{Fig:3} that can result in the field increasing with 
respect to the case of sphere with a smooth surface.

\textit{Conclusions}. -- The theory of the laser-assisted second harmonic
generation in non-spherical nanoparticles with free electrons was developed
and used to calculate the second harmonic both in the far-field and in the
near-field versus the laser frequency in subwavelength cubic metal
nanoparticles with steplike surface. The results have been compared with the
linear response field as well as with the same results for a spherical
nanoparticle. They show that the second harmonic in cubic nanoparticles is
generally much more intense as compared with the SHG in spherical 
nanoparticles of the same volume. The dominant resonance frequencies were
determined, which for the second harmonic correlate well with those for the
linear response case. It is also shown that for the near field the dominant
resonance frequencies can be different depending on the position near the
nanoparticle.

\end{document}